&
\documentstyle[11pt]{article}

\begin{document}

\title {\bf
Critical Behaviour of the Randomly Spin--Diluted 2-d Ising
Model --- A Grand Ensemble Approach\\}

\author{
     Reimer K\"uhn\thanks{Supported by a Heisenberg fellowship}
}

\date
      {
          Institut f\"ur Theoretische Physik   \\
                Universit\"at Heidelberg \\
                    Philosophenweg 19\\
             69120 Heidelberg, Germany\\
      }
  \vspace {3ex}

\maketitle

  \vspace {3ex}

\begin{abstract}
\noindent The critical behaviour of the randomly spin-diluted Ising model in
two space dimensions is investigated by a new method which combines a grand
ensemble approach to disordered systems proposed by Morita with the
phenomenological renormalization group scheme of Nightingale. Accurate
approximations for the phase diagram and for the connectivity length exponent
of the percolation transition are obtained. Our results suggest that the
thermal
phase transition of the disordered system might be different from that of the
pure system: we observe a continuous variation of critical exponents with the
density $\rho$ of magnetic impurities, respecting, however, weak universality
in the sense that $\eta$ and $\gamma/\nu$ do {\it not\/} depend on $\rho$ while
$\gamma$ and $\nu$ separately do. Our results are in qualitative and
quantitative agreement with a recent Monte--Carlo study.
\end{abstract}

\vfill\eject
The present contribution is concerned with the critical behaviour of the
randomly spin-diluted (SD) Ising model in two space dimensions. For this model,
the Harris criterion [1] according to which small amounts of disorder do (do
not) change the nature of the phase transition, if the corresponding pure
system's specific heat exponent is positive (negative), is inconclusive, since
$\alpha_{\rm pure} = 0$. The critical behaviour of the disordered system has
been a subject of debate for many years; see {\it e.g.} [2] and references
therein, and [3-10].

Currently, there appears to be widespread consensus --- based upon a
field--theoretic reformulation of the system with {\it weak\/} bond--disorder
(BD) in the critical regime --- that, in this limit, critical exponents of the
disordered system are the same as those of the pure system, albeit modified by
logarithmic corrections [3-6]. Numerical simulations of  BD systems [7,8]
indicate that such modifications through logarithmic corrections would persist
even in the strongly disordered regime. This consensus is, on the other hand,
being questioned in [9], where a more complicated phase transition, and a
non--divergent specific heat are predicted. Lastly, a very recent large--scale
Monte--Carlo study of the SD Ising system [10]  produces results which are at
variance with the findings of [3-9], in that it gives critical exponents which
clearly vary with the density $\rho$ of occupied sites.

It is this investigation, in particular, which finally convinced us to put our
own transfer--matrix analysis of the SD system --- both the method, and its
results --- to public discussion, even though, or rather {\it because\/} we are
aware that a deeper understanding of our approach [11] would certainly still be
welcome. In essence,  we find that there is full qualitative and quantitative
agreement between the Monte--Carlo results of Kim and Patrascioiu [10] and
ours.

We study the SD system by a new method [11] which combines Morita's grand
ensemble approach to disordered systems [12] with phenomenological
renormalization [13].
We begin by briefly describing Morita's method of configuration averaging in
systems with quenched randomness. Within this approach, a hierarchy of
approximating equilibrium systems is constructed, whose critical behaviour in
turn is obtained by phenomenological renormalization. We then state the main
results, relegating a detailed description of the method to a separate
publication.

Consider a disordered system described by the Hamiltonian $H(\sigma|\kappa)$,
where $\kappa$ denotes the quenched disorder configuration, and $\sigma$ the
set of dynamic variables of the system. Morita's approach avoids configuration
averaging of the free energy by working in an enlarged phase space in which
$\kappa$, too, is a dynamic variable, and by introducing a potential
$\phi(\kappa)$
chosen such that the new system with hamiltonian $H^\phi(\sigma,\kappa) =
H(\sigma|\kappa) + \phi(\kappa)$ exhibits thermodynamic equilibrium properties
identical to the non--equilibrium properties of the original quenched system.
To achieve this, the distribution $p^\phi(\sigma,\kappa)$ generated by
$H^\phi(\sigma,\kappa,)$ must be constructed such
that it satisfies
\begin{equation}
p^\phi(\sigma,\kappa) =\frac{1}{Z^\phi} \exp[-\beta H^\phi(\sigma,\kappa)] =
\frac{q(\kappa)}{Z(\kappa)} \exp[-\beta H(\sigma|\kappa)]
\end{equation}
for all $(\sigma,\kappa)$, or equivalently  $-\beta\phi(\kappa) = \ln
[q(\kappa)/ Z(\kappa)] + \ln Z^\phi$. Here $Z^\phi$ and $Z(\kappa)$ are
partition functions of the grand ensemble and the quenched system with fixed
impurity configuration $\kappa$, respectively; $q(\kappa)$ is the probability
distribution describing the quenched disorder. If $\phi$ is normalized such
that its configuration average vanishes [14], Eq. (1) implies
\begin{equation}
\ln Z^\phi = <\ln Z(\kappa)>_q - <\ln q(\kappa)>_q \ ,
\end{equation}
where $<..>_q$ denotes an average over the quenched disorder. That is, $\ln
Z^\phi$ gives the Brout free energy plus an irrelevant contribution of an
entropy of mixing.

Eqs. (1) and (2) are not very useful for practical calculations. In order to
utilize Morita's ideas, one has to find a representation for $\phi(\kappa)$
which is
adapted to the problem at hand, and one will usually have to resort to
approximations [11,15--17]. For the spin--diluted Ising model, one may expand
$\phi(\kappa)$ according to
\begin{equation}
\beta\phi(\kappa) = \lambda_0 + \lambda_1 \sum_i k_i + \lambda_2 \sum_{(i,j)}
k_i k_j + \dots + \lambda_P\sum_P \prod_{i\in P} k_i + \dots
\end{equation}
where the $k_i$ are occupation numbers taking the value 1 or 0 if in $\kappa$
the site $i$ is occupied or empty [18]. The first sum in (3) is over all
lattice sites, the second over nearest neighbour pairs, and the third over all
elementary plaquettes of the system. Each term serves to control one moment of
the probability distribution $p^\phi(\kappa) = \sum_\sigma
p^\phi(\sigma ,\kappa)$ - i.e. an expectation under $p^\phi$ of some product of
occupation numbers. The couplings $\lambda_1,\lambda_2,...$ have to be
determined as functions of temperature and field such that the moments of
$p^\phi(\kappa)$ coincide with those of $q(\kappa)$. The constant $\lambda_0$
is required to achieve $<\phi(\kappa)>_q=0$.

For a randomly spin-diluted system with density $\rho$ of magnetic impurities,
one would have to determine the couplings of $\phi$ so that
\begin{equation}
<k_i>_\phi=\rho\ ,\quad <k_i k_j>_\phi=\rho^2\ ,\dots ,\
<\prod_{i\in P} k_i>_\phi = \rho^{|P|}\ ,
\end{equation}
and so on [11,16]. Here $<..>_\phi$ denotes an average with respect to
$p^\phi(\kappa)$ and $|P|$ the size of the elementary plaquette, which for the
square lattice is 4. Eqs. (4) constitute an infinite set of equations for the
couplings $\lambda_1,\lambda_2,...$ of the potential $\phi$. To obtain a full
solution is, in general, impossible. Interpreting (4) as a set of constraints
imposed on the thermal motion of the magnetic impurities, one may, however, set
up a systematic scheme of approximations by letting only finite subsets of this
set of constraints become operative [11,16]. Implementing only $<k_i>_\phi =
\rho$ , one would describe an annealed system at density $\rho$ of magnetic
impurities. Such a system would condense and order at low temperatures and
would thus provide a rather poor description of the quenched system. If, in
addition,
one fixes nearest neighbour correlations $<k_i k_j>_\phi$ of particle locations
at their quenched value $\rho^2$, the condensation phenomenon no longer occurs
and the system exhibits, e.g., a percolation transition. Such a system is
already a serious candidate for the description of fully frozen--in disorder.
In this manner, one arrives at increasingly accurate descriptions of the
quenched
system as more and more constraints are taken into account, until eventually
one would obtain an exact description of the original disordered system. The
hope, of course, is that already rather simple approximations in this hierarchy
might belong to the ``univerality class" of the quenched system.

Except for the one dimensional system [11], this program cannot be carried
through to the end, and for $d \ge 2$ we know of no exact solution of even the
simplest approximating systems. At all levels of approximation, though, one has
to do with translationally invariant equilibrium systems. Their critical
behaviour could, for instance, be obtained by standard RG methods, were it not
for the fact that the couplings of $\phi$ are only determined through a set of
constraints. Since it is far from obvious how these constraints should be
transformed under rescaling (see, however, [16]), we decided to use
phenomenological renormalization [13], which avoids this problem altogether. No
explicit RG transformation in the space of couplings need be constructed. Given
$\rho$, one just solves the system in strip geometries - with the appropriate
set of constraints (4) imposed [11].

If $\xi_M(\rho,T)$ denotes the correlation length of an
$(M\times\infty)$--system at density $\rho$, temperature $T$, and field $H=0$,
an approximation to the critical point $T_c$ is given by the fixed point of the
phenomenological RG relation  $\xi_M(\rho,T)/M = \xi_{M'}(\rho,T')/M'$. An
approximation to the correlation length exponent $\nu$ may be obtained by
linearizing this relation about its fixed point [13]. Note that $\rho$ is a
parameter of this procedure. By varying $\rho$, we can obtain the phase
boundary $T_c(\rho)$, and the thermal correlation length exponent $\nu_t$ along
the
critical line. Analogous relations can be used to study the percolation
transition at $T = 0$, $\rho = \rho_c$.

We have studied four approximating systems, named (a) - (d). In system (a),
only
the first and second constraint displayed in (4) are imposed on the system. Due
to the anisotropy of the strip geometry, correlations parallel and
perpendicular to the strip turn out to be different. We have thus introduced
system (b) where
correlations parallel and perpendicular to the strip are treated as separate
constraints. Systems (c) and (d) are obtained from (a) and (b) by fixing, in
addition, correlations around each elementary plaquette at $\rho^4$. Note
that the results obtained from systems (a) and (b) or (c) and (d) should
approach each other as the strip widths $M$ and $M'$ go to infinity. This can
serve as a valuable consistency check for extrapolations of critical parameters
obtained from fixed points of the phenomenological RG relations to the infinite
system values.

We now turn to the results.
We have computed critical temperatures $T_c(\rho)$ for various densities $\rho
< 1$. The values obtained for systems (a) - (d) agree very well down to $\rho
\simeq 0.75$. Below  $\rho \simeq 0.75$, critical temperatures of the
approximating systems with the plaquette--constraint imposed turn out to be
slightly larger than those of systems (a) and (b) without this constraint.
Extrapolating a parabolic interpolation of the $T_c(\rho)$ values for $\rho=
0.9$, $\rho = 0.95$, and $\rho= 0.975$ to $\rho = 1$, we obtain $T_c(\rho) \to
2.2674$, as $\rho \to 1$, which is off the
mark by less than $0.1\%$, and $T_c(\rho)^{-1} d T_c(\rho) / d\rho \to 1.579$,
as  $\rho \to 1$, which differs from the exact result 1.535 [19] by less than
$3\%$. We have also checked the scaling form of the transition line near the
percolation transition at $\rho_c$, $\exp[-2J/k_B T_c(\rho)]
\sim (\rho - \rho_c)^\varphi$, with $\varphi = \nu_t (\rho_c)/ \nu_p$, and we
find $\varphi = 1$ to within less than $2\%$ for all four approximating systems
[11]. Here $\nu_p$ denotes the connectivity--length exponent of the percolation
transition. The percolation threshold $\rho_c$ itself ($\rho_c \simeq 0.593$
[20]) is correctly reproduced to within $3\%$ and $1\%$ by systems (a),(b) and
(c),(d) respectively; see table I.

Fig. 1 shows our results for the correlation length exponent $\nu$ of system
(a), based on extrapolations from strip widths up to $M=8$ and $M=10$ for the
thermal and the percolation transitions, respectively. They clearly show a
variation with the density $\rho$ of magnetic impurities. For $\rho=0.7$ and
$\rho=0.9$, table I shows that this variation persists, as we impose further
constraints in approximating systems (b)--(d). Note that the correlation length
exponents are always mutually consistent for these four approximating systems.
For the percolation transition, they agree well with the expected exact result
$\nu_p = 4/3$ of den Nijs [26]. Also, the isotropic and anisotropic variants in
the two groups of systems, {(a),(b)} and {(c),(d)}, always give consistent
results (as they should). While all four systems agree in ``universal"
characteristics of the phase transition, $\nu_p$ or $\nu_t$, the first group
can differ from the second as far as non-universal parameters of the phase
transitions, i.e. $\rho_c$ or $T_c(\rho)$, are concerned.

\begin{table}[h]
{ \caption{ \label{Tab1}  \protect\small
Extrapolated critical parameters for the percolation transition, and
the thermal phase transition at  densities $\rho  = 0.7$ and $\rho = 0.9$ of
magnetic impurities. Numbers in brackets give the estimated error of the last
displayed digit of the preceding quantity.}}

                                                          \begin{center}
\begin{tabular}{|l|l|l|l|l|l|l|}
\hline
\hline
&\multicolumn{2}{c|}{percolation}&\multicolumn{2}{c|}{$\rho = 0.7$} &
\multicolumn{2}{c|}{$\rho = 0.9$}\\
&\multicolumn{1}{c}{$\rho_c$} &\multicolumn{1}{c|}{$\nu_p$}
&\multicolumn{1}{c}{$T_c$}    &\multicolumn{1}{c|}{$\nu_t$ }
&\multicolumn{1}{c}{$T_c$}    &\multicolumn{1}{c|}{$\nu_t$}\\
\hline
                                                              (a) & 0.609(1) &
1.33(2) & 1.051(1) & 1.31(2) & 1.902(1) & 1.13(1)\\
(b) & 0.609(1) & 1.33(2) & 1.050(1) & 1.30(2) & 1.901(1) & 1.13(1)\\
(c) & 0.587(1) & 1.34(2) & 1.081(1) & 1.29(2) & 1.901(1) & 1.12(1)\\
(d) & 0.587(1) & 1.33(2) & 1.080(1) & 1.30(2) & 1.901(1) & 1.13(1)\\
\hline
\hline
\end{tabular}
\end{center}
\end{table}

The variation of critical exponents with the density $\rho$ is
at variance with what would be expected from universality arguments, and one
might well argue it to be an artefact of our simple approximations. So
independent evidence is clearly welcome. Therefore, we turned to standard
real--space renormalization group (RSRG) calculations for this system [2], and
noted that they, too, {\it are}\/ able to produce similar results: Some time
ago, Tsallis and Levy [21] used the RSRG to compute phase diagrams of bond
diluted models, and they propose, {\it inter alia}, a set of RSRG
transformations {\it parametrized\/} by the density $\rho$ of bonds. Applying
this idea to the SD problem, and using techniques of [22], we obtain the full
curves shown in Fig. 1. While such an approach may still seem {\it ad hoc}, we
find that, last but not least, very recent large scale Monte--Carlo simulations
of the SD system [10] also give a non--universal $\nu_t(\rho)$ in {\it
quantitative\/} accord with our results, as can be seen in Fig. 1.

The agreement between the results of Kim and Patrascioiu [10] and ours goes, in
fact, much deeper: We computed susceptibilities and used the finite--size
scaling relation $\chi_M(T_c) \sim M^{\gamma/\nu}$ to determine $\gamma/\nu$.
We find $\gamma/\nu \simeq 1.75$ as in the pure system, and {\it independently}
of $\rho$, which implies that $\gamma$ itself is again non--universal. The
amplitude $A_0$ of the critical finite--size correlation length, $\xi_M(T_c)
\simeq A_0 M$, is known to be related to the critical exponent $\eta$ according
to $A_0 = 1/\pi\eta$ [23]. We have used this relation to determine $\eta$, and
we find $\eta \simeq 0.25$ independently of $\rho$ for the thermal transition,
which is of course consistent with the $\rho$--independence of $\gamma/\nu$ via
the Fisher relation $\gamma/\nu = 2 - \eta$. We find this relation satisfied by
all our extrapolated values for all our approximating systems, usually to
within
$1\%$ or better. The same kind of weak universality [24] is also observed in
[10], and it is of the same type as that encountered in Baxter's 8--vertex
model [25].

At this point let us mention that some time ago Derrida {\it et al.} [27] found
analogous non-universal behaviour of Binder's cumulant ratio in a family of
self--dual bond disordered Ising models, which they ascribed to logarithmic
corrections to finite--size scaling. However, as pointed out by Cardy [28], in
order to disentangle logarithmic from power-law corrections to finite-size
scaling, one might have to go to rather large strip widths, and our data, as
yet, do not support the conclusion that the observed non--universality is only
apparent and caused by logarithmic corrections to finite--size scaling; see
Fig. 2 which shows the variation of the phenomenological critical exponent
$\nu_{M,M-1}$ of system (a) with system size $M$, and the insert which adresses
this particular point.

The complete set of constraints (4) imposes the condition $<\prod_{i\in
\omega}\ k_i>_\phi = \rho^{|\omega|}$ for all subsets $\omega$ of the lattice,
implying
that occupancy of lattice sites is uncorrelated, or in other words that the
correlation length describing the $k_i$--correlation functions vanishes. Our
simple approximations only fix some of these correlations. Others can of course
be computed and are found to vary only slightly with  temperature and field. We
have checked that the corresponding correlation lengths remain small (at most a
few lattice spacings) so that the approximating systems may be regarded as
virtually quenched. Moreover, it can be shown [11] that in zero external field
an infinite set of couplings of the exact potential $\phi$ vanishes, and is
thus
correctly taken into account already at the level of our simple approximating
systems (It implies, for instance, that in zero field our system (a) already
provides an exact description of the quenched system in 1-d).

In summary, we have studied the critical behaviour of the spin--diluted Ising
model by a new method which combines a grand ensemble approach to disordered
systems with phenomenological renormalization. Where a comparison with exact
results (concerning the phase diagram but also universal aspects of the
percolation transition) is possible, our data compare favourably with earlier
RG analyses of the system (see e.g. [2] and Refs. therein). We know of no
a-priori
reason why the present method should be intrinsically better in describing the
percolation transition, where we get a value of $\nu_p$ consistent with the
exact result of den Nijs, than in describing the thermal phase transition at
$\rho_c < \rho \le 1$, where our results indicate a continuous variation of
critical exponents with $\rho$ --- in a manner that respects weak universality.
Indeed, a recent Monte--Carlo study [10] produces results completely in accord
with ours, both qualitatively and quantitatively. Hence, the the observed
variation of critical exponents is most likely {\it not} an artefact of our
grand ensemble description of quenched disorder.

A deeper understanding of our method would certainly still be welcome. Let us
mention in closing that it is easily adapted to the study of correlated
disorder. Preliminary results for a 1-d system, where comparison with an exact
solution is possible [29] do look encouraging. We are currently also using our
method to study the wetting transition in the presence of surface disorder,
which has recently been a subject of some controversy in the literature.

{\bf Acknowledgements}
This work was partly supported by the Deutsche Forschungsgemeinschaft. I am
indebted to A. Huber for having acquainted me with Morita's ideas and for
helpful advice at an earlier stage of this project. Illuminating discussions
with B. Derrida, J. Vannimenus, and F. Wegner are also gratefully acknowledged.

\bigskip
\bigskip
\noindent
{\bf Figure captions}\\

\noindent
{\bf Fig. 1:} Correlation length exponent $\nu$ of system (a) for various
densities
$\rho$ (open squares). The leftmost open square was determined as a
connectivity
length exponent of the percolation transition. The exact result $\nu_p = 4/3$
is displayed as a diamond at $\rho_c\simeq 0.593$. Results from a RSRG
calculation, using the b=2 and b=3 decimation transformations of Yeomans and
Stinchcombe are given as full curves. Full circles show the Monte--Carlo data
of [10]; except near the percolation transition, where a precise location of
the phase boundary is difficult, the agreement with our model (a) results is
excellent.\\

\noindent
{\bf Fig. 2:} Variation of the phenomeneological critical exponent
$\nu_{M,M-1}$ with strip width $M$, for approximating system (a); open squares:
percolation;
diamonds: $\rho = 0.7$, full squares: $\rho = 0.9$, open circles: pure
system. The insert checks for the possibility of logarithmic corrections to the
pure systems critical behaviour: The quantity $1/\Delta_M  = [\nu_{M,M-1}
(\rho) - \nu_{M,M-1}(1)]^{-1}$ is plotted vs. $\ln (M)$. For $\rho = 0.9$,
$1/\Delta_M$ is not monotonically increasing with $M$ and levels off for large
$M$, which is evidence against logarithmic corrections. For $\rho = 0.7$  the
data themselves are not as conclusive as for $\rho = 0.9$. However,
$1/\Delta_M$ appears to increase slower than linearly with $\ln (M)$ and there
is a trend for the curve to level off.


\begin{thebibliography}{88}
\bibitem [18] {kk} An important technichal point deserves mention here. To
avoid
counting problems in the grand ensemble, we eventually lump the products $k_i
\sigma_i$ appearing in the Hamiltonian of the SD Ising model into a spin--one
variable $S_i \in \{0,\pm 1\}$, so $k_i =|S_i|$
\end{thebibliography}
\end{document}